\documentclass[11pt]{article}

\usepackage[english]{babel}
\usepackage[utf8]{inputenc}

\usepackage{amssymb,dsfont,amsmath,amsfonts,hyperref,mdframed,mathrsfs,stmaryrd,bbold}
\usepackage[enableskew]{youngtab}
\usepackage[active]{srcltx}
\usepackage{slashed}
\usepackage{color}
\advance \textheight by \topskip
\DeclareMathAlphabet{\mathpzc}{OT1}{pzc}{m}{it}
\usepackage{xcolor}

\usepackage[normalem]{ulem}
\usepackage{graphicx}
\usepackage{soul}
\usepackage{cite}

\usepackage{mathtools}

\usepackage{array}

\usepackage{cancel}

\definecolor{mygreen}{rgb}{0.1, 0.7, 0.2}

\def\be{\begin{equation}}
\def\ee{\end{equation}}
\def\bea{\begin{eqnarray}}
\def\eea{\end{eqnarray}}

\def\eg{\textit{e.g.}}

\def \half{{spin--$\tfrac12$}}

\def\part{\partial}

\begin{document}

\title{Quantization of counterexamples to Dirac's conjecture}

\author{
Mauricio Valenzuela\\[12pt]
\textit{Facultad de Ingenier\'ia, Arquitectura y Dise\~no}, \textit{Universidad San Sebasti\'an, Valdivia, Chile}\\
\textit{Centro de Estudios Cient\'{\i}ficos (CECs), Arturo Prat 514, Valdivia, Chile } \\
}

\maketitle

\begin{abstract}

Dirac's conjecture, that secondary first-class constraints generate transformations that do not change the physical system's state, has various counterexamples. Since no matching gauge conditions can be imposed, the Dirac bracket cannot be defined, and restricting the phase space first and then quantizing is an inconsistent procedure. The latter observation has discouraged the study of systems of this kind more profoundly, while Dirac's conjecture is assumed generally valid. We point out, however, that secondary first-class constraints are just initial conditions that do not imply Poisson's bracket modification, and we carry out the quantization successfully by imposing these constraints on the initial state of the wave function. We apply the method to two Dirac's conjecture counterexamples, including Cawley's iconical system.

\end{abstract}

\tableofcontents

%%%%%%%%%%%%%%%%%%%%%%%%%%%%%%%%%%%
\section{Introduction}
%%%%%%%%%%%%%%%%%%%%%%%%%%%%%%%%%%%

In gauge systems, passing from the Lagrangian to the Hamiltonian formulation, the Legendre transform produces \textit{primary constraints} on the phase space. These primary constraints alone are sufficient to reverse the Legendre transform and recover the Lagrangian dynamics through the Hamiltonian least action principle, accounting for gauge redundancies. Adding the canonical Hamiltonian and the primary constraints with their respective Lagrange multipliers defines the total Hamiltonian, $H_T$, which reproduces the field equations using Poisson brackets.

 In \textit{Lecture on quantum mechanics} \cite{Dirac}, Dirac stated that:
``I think it may be that all the first-class secondary constraints should be included among the transformations which don't change the
physical state, but I haven't been able to prove it. Also, I haven't found any example for which there exists first-class secondary constraints which do generate a change in the physical state." 

Subsequently, Dirac proposed modifying the field equations to promote the change produced by secondary first-class constraints to true gauge symmetries. This is done by adding the secondary first-class constraints to the total Hamiltonian, which defines Dirac's extended Hamiltonian $H_E$. The evolution generated by $H_E$ incorporates arbitrary velocities given by the first-class Lagrange multipliers, for variables conjugate to secondary first-class constraints, rendering them pure-gauge degrees of freedom. 

However, various counterexamples of Dirac's conjecture have been found. The states in the orbits generated by secondary first-class constraints are not redundant or unphysical. Cawley \cite{Cawley} provided the first counterexample, also discussed in \eg \cite{Frenkel:1980nt,Gomis:1986mn,Montani:1998ip,deLeon:2020wdu,Tomonari:2023vgg}, and others can be found in \cite{Henneaux:1992ig,Li_1993,Valenzuela:2022zic,Jinno:2020zzs}. For discussions on the general validity of Dirac's conjecture, see \cite{Castellani,Gotay,Gracia:1988xp,Earman:2003cy,Pons:2004pp}.

In these cases, Dirac's extended dynamic would lead to loss of physical degrees of freedom, demoted to pure-gauge variables instead. Still, Dirac's conjecture is considered generally valid. One possible reason for this can be found in \cite{Henneaux:1992ig}, where it is argued that the subspace of the phase space defined by the secondary first-class constraints that are not accompanied by gauge fixing conditions, has no bracket structure, which prevents quantization. In this paper, we find a way out of this obstruction.

We shall point out that secondary constraints are just initial conditions: since they do not contain time derivatives, they are equivalent to restrictions on the initial Cauchy surface. Hence, imposing these constraints as initial conditions can be done without changing the bracket structure, and they can be quantized as operators annihilating the initial state of the wave function.

We apply this technique to two Dirac's conjecture counterexamples, including Cawley's system \cite{Cawley}, and their quantization is thus achieved. 

The organization of this paper is as follows: In Section \ref{sec:HT}, the elements of Hamiltonian analysis necessary for this article are introduced. We describe first Dirac's conjecture counterexamples in Section \ref{sec:count1}, the second counterexample corresponds to Cawley's systems given in Section \ref{sec:count2}. Finally, our conclusions are presented in Section \ref{sec:conc}.

\section{Dynamics from the total Hamiltonian}\label{sec:HT}\

Given a mechanical system, passing from the Lagrangian formulation in configuration space $\mathcal{Q}$, to the phase space $\mathcal{P}$, involves the Legendre transform,
\be\label{Leg}
p_s=\frac{\partial L}{\partial\dot q^s}.
\ee
Thus, coordinates and velocities $(q^s,\dot q^s)$, labeled by $s$, are mapped to position and momenta variables $(q^s,p_s)$ labeling points in  $\mathcal{P}$. The extended phase space, with points $(t,q^s,p_s)$ will be denoted  $\mathcal{T}$. 
Then from the Lagrangian $L$, we obtain the canonical Hamiltonian,
$$
H_C=\dot q^s p_s -L\,.
$$

The phase space comes with the (fixed time) Poisson bracket,
\bea
\{f(q,p),g(q,p)\}&=&  \sum_s \left(\frac{\partial f(q,p)}{\partial q^s(t)} \frac{\partial g(q,p)}{\partial p_s(t)}- \frac{\partial f(q,p)}{\partial p_s(t)}\frac{\partial g(q,p)}{\partial q^s(t)} \right)\,.
\eea

If relations between positions and momenta appear in \eqref{Leg}, they are expressed in homogeneous form, $c_{(0)}(q,p)\approx0$, and dubbed \textit{primary} constraints. The primary constraints can be \textit{first-class}, \textit{second-class}, or a combination of them. 

The Hamiltonian action functional, 
\be\label{ST}
S_T=\int dt (\dot q^s p_s-H_T)
\ee
where 
\be
H_T:=H_C+  c_{(0)} \cdot \mu_{(0)}\,,
\ee
the \textit{total Hamiltonian}, incorporate the scalar product of primary constraints, $c_{(0)}$ is an array fo them, times Lagrange multipliers $\mu_{(0)}$. Adding primary constraints to the Hamiltonian is necessary to reproduce the equivalent of the Euler-Lagrange equations from the least action principle \eqref{ST}. 

The field equations can also be obtained from
\be
\dot f= \{f,H_T\}\,,\qquad c_{(0)}(q,p)\approx0\,. \label{HTeqc}
\ee
The Poisson bracket part will not reproduce the primary constraints $c_{(0)}\approx 0$ alone, and we should recall them by putting them aside.

The primary constraints $c_{(0)}(q,p)\approx 0$ remind us that the system can only evolve in a subspace of the phase space, upon which the Lagrangian system is mapped. The directions outside the region described by $c_{(0)}(q,p)\approx 0$ are never reached as mapped from trajectories in the original Lagrangian system. Hence no velocities will be developed outsides those directions, and we have,
\be \label{dtn}
\frac{d^nc_{(0)}}{dt^n}= 0\,,\qquad n\geq1 \,.
\ee
Hence and from the first part of the field equations \eqref{HTeqc}, we obtain 
\be\label{nary}
c_{(n)}(q(t),p(t)):=\{H_T,c_{(n-1)}\}=0 \,, \qquad n\geq 1 \,,
\ee
denoted secondary constraints. Since $\frac{d^nc_{(n)}}{dt^n}=0$, these equations determine initial conditions. This observation is valid for secondary constraints of generic systems, including the Gauss law in electromagnetism (See p. 345 \cite{wein-qft}).

The presence of first-class constraints among $c_{(0)}(q,p)\approx 0$ is a source of indeterminism since the Lagrange multipliers of primary first-class constraints remain arbitrary, in contrast with second-class constraints Lagrange multipliers which are determined by consistency conditions.

Indeed, assuming for simplicity that the primary constraints are first-class, the time-evolution indeterminacy relies on the arbitrary velocities produced by the Lagrange multipliers $\mu_{(0)}$, associated with primary first-class constraints, since
\be
\dot f= \{f,H_C\}+ \{f, c_{(0)}\} \mu_{(0)}\,,
\ee
on variables such that $\{f, c_{(0)}\}\neq 0$. For $c_{(0)}$ linear functions of the phase space coordinates, the variables developing arbitrary velocities are those which are conjugate to $c_{(0)}$. For example, in electromagnetism, using the standard notation, $c_{(0)}=\pi^ 0$ and its conjugate variable is $A_0$. Hence $A_0$ has arbitrary evolution since $\dot A_0 = \mu$ is a Lagrange multiplier. Even though specific initial conditions are given for $A_0$, its value in the future is indeterminate.

Systems with first-class constraint possess gauge symmetries, which are generated by the Castellani chain \cite{Castellani}, 
\be\label{chain}
G=\sum_{n=0}^k \epsilon^{(k-n)}G_{n}\,,\qquad \epsilon^{(l)}:=\frac{d^{l}\epsilon}{dt^{l}}\,,
\ee
a sequence of first-class constraints $G_{n}$ times derivatives of the gauge parameter. Assuming that there are no second-class constraints among $c_{(n)}$ we would have $c_{(n)}=G_{n}$.

We will also need the following definition. Secondary constraints are denoted \textit{ineffective} when the line elements,
\be
dc_{(n)}(q(t),p(t))\approx 0\,,\qquad d:=dq^s\frac{\partial}{\partial q^s}+dp^s\frac{\partial}{\partial p^s}\,,
\ee
vanishes on the surface of the constraints $c_{(...)}=0$, otherwise they are called \textit{effective}. The ineffective constraints do not generate transformations.

\section{A counterexample to Dirac's conjecture}\label{sec:count1}

The following Dirac's conjecture counterexample can be found in \cite{Henneaux:1992ig}.
The Lagrangian is given by
\be\label{L1}
L= \frac12 \exp(y) \dot x^2.
\ee
And the Euler-Lagrange equations read,
\be\label{ELce1}
e^y \dot x^2=0\,,\qquad e^y( \dot y \dot x+\ddot x)=0\,.
\ee
discarding the $y=-\infty$ formal solution, the system can be reduced to the space of solutions described by $\dot x(t)=0$. Then $\ddot x(t)=0$ and $y(t)$ is indeterminite.

Passing from configuration space to the phase space variables, we obtain the velocity-momentum correspondence, 
\be\label{pxpy}
p_x= e^y \dot x \,,\qquad p_y\approx0\,,
\ee
which yields the primary constraint $c_{(0)}=p_y\approx0$. 

The total Hamiltonian reads, 
\be
H_T=\frac12 e^{-y} p_x^2+\mu p_y,\qquad p_y \approx 0\,.
\ee
Now the variation of the action \eqref{ST} yields the system
\be\label{Heqc1}
\dot x = e^{-y}  p_x,\qquad \dot y = \mu,\qquad \dot p_x =0,\qquad \dot p_y=\frac{e^{-y}}2 p^2_x,\qquad p_y \approx 0\,.
\ee
As a consistency condition, $\dot p_y\approx 0$ implies the secondary constraint
\be
c_{(1)}:=\frac{e^{-y}}2 p^2_x=0\,,
\ee
which implies $p_x=0$. Hence $\dot x=0$, and \eqref{Heqc1} reduces to, 
\be\label{Heqc1b}
\dot x =0,\qquad \dot y = \mu,\qquad \dot p_x =0,\qquad \dot p_y=0,\qquad p_x =0\,,\qquad p_y \approx 0\,,
\ee
and we recover the Euler-Lagrange equations \eqref{ELce1}, taking into account \eqref{pxpy}, and considering that $\mu$ is arbitrary. 

The equations of motion \eqref{Heqc1} can be obtained from the system \eqref{HTeqc},  
\be
\dot f = \{f,H_T\},\qquad p_y\approx 0\,.
\ee

The constraint $c_{(1)}$ holds if $p_x = 0$, hence it is innefective since  $\partial_x c_{(1)}\approx 0\approx \partial_{y} c_{(1)}$. Since $\dot p_x =0$, $p_x(t)=0$ determines the initial condition.
This implies that the gauge symmetry generator chain \eqref{chain} reduces to $G=\epsilon c_{(0)}$, and the field equations \eqref{Heqc1b} are invariant under transformations $\delta y = \epsilon(t)$,  $\delta \mu=\dot \epsilon$. 

The constraint $c_{(1)}$ generates no transformation, and its square root $p_x\approx 0$ is not a symmetry generator. Hence the Dirac conjecture does not hold.

The state of the system, $(x=x_0, p_x=0,y(t), p_y\approx0)$ is not yet determined, since the solution of the field equation $\dot y= \mu$ depends on the initial condition $y(0)$ and the undetermined function $\mu$.

\subsection{Fixing the gauge}

To break the time evolution indeterminacy, we fix the gauge imposing an algebraic constraint on the phase space, expressing $y$ as a function of the remaining phase space variables. In this case, the only non-trivial variable is $x$. We choose
\be\label{gf}
y-a(x)\approx 0\,,
\ee
for a given function $a(x)$. Since  $\{y(t)-a(x)\approx 0,p_y\approx0\}=1$ follows that $p_y\approx0$ and \eqref{gf} are second-class. The consistency condition $\dot y(t)-\dot a(x)\approx 0$ determines the Lagrange multiplier,
\be\label{mu}
\mu= \dot x \partial_x a=0\,,
\ee
from the field equation $\dot x =0$. Now the state of the system is specified by $(x=x_0, p_x=0, a(x_0), p_y\approx0)$, where $x_0=x(0)$ is an initial condition, and there is no time evolution arbitrariness.

The Dirac bracket, defined from $y(t)-a(x)\approx 0$ and $p_y\approx0$ second class-constraints, reduces to,
\be\label{D1}
\{f(x,p_x),g(x,p_x)\}_D= \frac{\partial f}{\partial x}\frac{\partial g}{\partial p_x} -\frac{\partial f}{\partial p_x}\frac{\partial g}{\partial x}\,,
\ee
on functions of $(x,p_x)$.

The gauge-fixed Hamiltonian reads,
\be
H_f=\frac12 e^{-a(x)}p_x^2.
\ee
The field equations are imposed with fixed initial condition $p_x=0$,
\be\label{Heq1}
\dot f=\{f,H_f\}_D\,,
\qquad p_x(0)=0\,.
\ee
and we obtain,
\be
\dot x=p_x,\qquad \dot p_x=0\,,\qquad p_x(0)=0\,,
\ee
with solutions,
\be
x(t)=x_0,\qquad p_x(0)=0.
\ee

\subsection{Quantization}

Projecting onto the surface of the second-class constraints $y(t)-a(x)\approx 0$ and $p_y\approx0$ defines a subspace of $\mathcal{P}$ that it is also phase space. This is Maskawa-Nakajima theorem  \cite{Maskawa:1976hw,wein-qft}.  There the Poisson structure takes the usual form in terms of coordinates $(x,p_x)$. 

If in addition the secondary constraint $p_x= 0$ is regarded as an algebraic constraint, as proposed in  \cite{Henneaux:1992ig}, we would have a projection from a two-dimensional phase space to the line, $(x, p_x)\rightarrow x$, which has no Poisson bracket, and hence the quantization would be impossible. To prevent this issue it is proposed to adopt the Dirac conjecture as a general principle. Then, treating $p_x\approx0$ as a gauge generator, the conjugate variable $x$ can be gauged away,  the constrained phase space is zero-dimensional, there are no degrees of freedom, and quantization is straightforward. 

We shall see that the latter approach can be avoided, in order to successfully quantize the system without loss of degrees of freedom. 

Since $\dot p_x=\{p_x,H_T\}=0$, $p_x(t)=0$ is equivalent to the initial condition $p_x(0)= 0$, the Poisson bracket defined on $(x,p_x)$ does not need to be changed. The secondary constraint can be reduced to restricting the initial Cauchy data, and $(x(0)=x_0, p_x(0)=0,y(0)\approx0, p_y(0)\approx0)$, together with the field equations \eqref{Heq1} fully determines the evolution of the system. The space of solutions consists of the trajectory $(x(t)=x_0, p_x(0)=0)$. 

The system is easily quantized promoting the Dirac bracket \eqref{D1} to the commutator and the Hamiltonian to the operator,
\be\label{hHf}
\hat H_f=\frac12 \mathcal{O} \widehat{(e^{-\hat a(x)} \hat p_x^2)},
\ee
where  $\mathcal{O}$ is the operator-ordering prescription. 

In the Schr\"odinger realization the momentum operator is given by $\hat p_x=-i\hbar \partial_x$,  and we can choose the hermitian form,
\be
\hat H_f=\frac14 \left(e^{-a(x)} \hat p_x^2+\hat p_x^2 e^{-a^*(x)}\right) \,.
\ee
Thus the Schr\"odinger equation reads, 
 \be
(i\hbar \partial_t-\hat H_f)|\psi(t)\rangle_{\hbox{\scriptsize{phys}}}
=0\,,\qquad \hat p_x|\psi(0)\rangle_{\hbox{\scriptsize{phys}}}=0\,,
\ee
where the condition on the initial state of the wave function is the quantum analog of the classical secondary constraint $p_x(0)=0$.

For $a(x)$ a linear function of $x$, the system's energy vanishes, and the wave function is a complex constant. 
The physical state has no energy, and its evolution is trivial $\partial_t |\psi(0)\rangle_{\hbox{\scriptsize{phys}}}=0$. It follows that $|\psi(0)\rangle_{\hbox{\scriptsize{phys}}}$ is stationary, and it remains true for all $t$,
\be\label{phys}
\hat p_x|\psi(t)\rangle_{\hbox{\scriptsize{phys}}}=0\,.
\ee
It is clear that the precise form of $a(x)$, as long as it remains linear in $x$, does not affect the physical state. For polynomial values of $a(x)$, we may have Gribov ambiguities, which we shall avoid here, and the Hamiltonian can acquire a non-vanishing eigenvalue.

\section{Cawley's counterexample}\label{sec:count2}

For the set of variables $(x,y,z)$, the Lagrangian reads,
\be
L=\dot x\dot z + \frac 12 y z^2,
\ee
from the definition of the momenta 
\be
p_x= \dot z \,,\qquad p_z= \dot x \,,\qquad p_y\approx 0\,,
\ee
we obtain the primary constraint $c_{(0)}=p_y\approx 0$. 

The total Hamiltonian is given by,
\bea
&H_T= p_x p_z-\frac12 y z^2+\mu p_y,&\\
&p_y \approx 0\,,&\label{py0}
\eea
and the equations of motion \eqref{HTeqc} yields,
\bea
&\dot x = p_z\,,\qquad \quad \dot y = \mu\,,\qquad \quad \dot z= p_x\,,&\label{Ceom1}\\
&\dot p_x =0\,,\qquad \dot p_y = \frac12 z^2\,,\qquad \dot p_z=yz\,.&\label{Ceom2}
\eea
From \eqref{py0}, $\dot p_y=0$ and \eqref{Ceom2}, we obtain the secondary constraint $c_{(1)}=-\frac12 z^2=0$, and the tertiary constraints $c_{(2)}=- z\dot z=-zp_x$, both ineffective. The root $z=0$ of $c_{(1)}$ implies $\dot z=p_x=0$, and \eqref{Ceom1}-\eqref{Ceom2} reduce to,
\bea
&\dot x = p_z\,,\qquad \qquad \dot y = \mu\,,\qquad \qquad \dot z=0&\label{Ceom3}\\
&p_x=0\,,\quad \dot p_x =0\,,\qquad p_y\approx 0\,,\quad \dot p_y =0\,,\qquad   z=0\,,\quad \dot p_z =0\,.&\label{Ceom4}
\eea

The constraints $(p_x,p_y,z)\approx 0$ are first-class. Hence assuming the Dirac conjecture, they would imply that $(x,y,p_z)$ are pure gauge, and there would be no physical degrees of freedom in the system. However, neither $c_{(1)}$, $c_{(2)}$, nor the roots $z\approx0$ and $p_x\approx0$ are gauge symmetry generators. The Castellani chain is reduced to $G_0=c_{(0)}$ and the gauge symmetry of the system is given by  $\delta y = \epsilon(t)$, $\delta \mu=\dot \epsilon$. 

Since $\dot z=0=\dot p_x$, secondary constraints are equivalent to the initial conditions, 
\eqref{Ceom1}-\eqref{Ceom2},
 \be\label{zpxt0}
z(0)\approx 0\,,\qquad p_x(0)\approx 0\,.
\ee
Indeed, the iteration of the time derivatives in the field equations yields,
\be\label{znpn}
\frac{d^nz}{dt^n}(0)=0\,,\qquad \frac{d^np_x}{dt^n}(0)=0\quad  \forall \quad n\,,
\ee
which imply $z(t)= 0$ and $p_x(t)=0$, for all $t$, from the Taylor expansions around $t=0$.

\subsection{Fixing the gauge}

Cawley's system's independent arbitrary functions of time are $\mu$ and $y$. Fixing the gauge implies choosing $y=a(x,p_z)$ as a linear function of the remaining non-vanishing phase space coordinates. The constraints $(y-a(x,p_z),p_y)\approx0$ form a second-class system, and hence $y$ and $p_y$ can be removed. The gauge-fixed Hamiltonian reads,
\bea
H_f= p_x p_z-\frac12 a(x,p_z) z^2,
\eea
and the field equations are imposed together with initial conditions,
\be\label{Heq2}
\dot f=\{f,H_f\}_D\,, \qquad z(0)=0,\, \qquad p_x(0)=0\,,
\ee
where the Dirac bracket reduces to,
\be\label{D1}
\{f,g\}_D= \frac{\partial f}{\partial x}\frac{\partial g}{\partial p_x} -\frac{\partial f}{\partial p_x}\frac{\partial g}{\partial x}+ \frac{\partial f}{\partial z}\frac{\partial g}{\partial p_z} -\frac{\partial f}{\partial p_z}\frac{\partial g}{\partial z}\,,
\ee
on functions of $(x,p_x,z,p_z)$.
Hence we obtain the field equations,
\bea
&\dot x = p_z\,,\qquad \dot z= p_x-\frac 12 \frac{\partial a}{\partial p_z } z^ 2\,,&\label{Ceom1f}\\[5pt]
&\dot p_x =\frac 12 \frac{\partial a}{\partial x }z^2\,,\qquad \dot p_z=az-\frac 12 \frac{\partial a}{\partial p_z }z^2\,.&\label{Ceom2f}
\eea
From this system, we can iterate time derivatives to find \eqref{znpn}, which implies $z(t)=0$ and $p_x(t)=0$ for all $t$ and \eqref{Ceom1f}-\eqref{Ceom2f} reduce to,
\bea
&\dot x = p_z\,,\qquad  \dot p_z=0\,.&
\eea
Hence the solution of the system is given by,
\be\label{sol}
 x(t) = t p_{0z}+x_0\,, \qquad z(t)=0\,,\qquad p_x(t)=0\,,\qquad  p_z(t)=p_{0z}\,,
 \ee
where $p_{0z}$ and $x_0$ are initial values not determined by the field equations.

The solutions \eqref{sol} are solutions of the original system \eqref{Ceom1}-\eqref{Ceom2} on the surface of the gauge fixing conditions.

\subsection{Quantization}

In the Schrodinger realization,
\be
\hat p_x=-i\hbar \partial_x\,,\qquad \hat z=i\hbar \partial_{p_z}\, \hat x=x\,,\qquad \hat p_z=p_z\,
\ee
the gauge-fixed Hamiltonian quantum analog is given by, 
\be
\hat H_f=  p_z \hat p_x -\frac14 (a(x,p_z) \hat z^2 + \hat z^2 a^\dagger(x,p_z))\,,
\ee
which is hermitian. 

Preparing the wave function in an initial state such that the quantum analogs of \eqref{zpxt0} holds, the Schrodinger equation reads, 
\be
(i\hbar \partial_t-\hat H_f)|\psi(t)\rangle_{\hbox{\scriptsize{phys}}}=0\,,\qquad \hat p_x|\psi(0)\rangle_{\hbox{\scriptsize{phys}}}\,,\quad \hat z|\psi(0)\rangle_{\hbox{\scriptsize{phys}}}=0\,.
\ee
Hence the initial state possesses no energy and it remains stationary,
\be
 |\psi(t)\rangle_{\hbox{\scriptsize{phys}}}=|\psi(0)\rangle_{\hbox{\scriptsize{phys}}}\,.
 \ee
As before, we have assumed that $a(x,p_z)$ is linear.

Note that the classical solution \eqref{sol} propagates as a free particle, however, the classical and quantum Hamiltonian vanish for the solution of the system. 

\section{Conclusions}\label{sec:conc}

Dirac's logical framework assumes that the constraints' primary or secondary nature is not physically relevant. However, Dirac's conjecture counterexamples suggest that their different nature must be recalled. Otherwise, considering primary and secondary first-class constraints on equal footing would lead us to eliminate physical degrees of freedom. 

Indeed, a relevant counterexample is the case of the massless Rarita-Schwinger system considered in supergravity \cite{fnf,dz}.
In references \cite{Valenzuela:2022gbk,Valenzuela:2023aoa} (see also \cite{Valenzuela:2023jcn}) it was shown that the elimination of the \half ~degree of freedom in \cite{Deser:1977ur,Senjanovic:1977vr,Pilati:1977ht,Fradkin:1977wv} was a consequence of Dirac's conjecture assumption, which otherwise propagates \cite{Alvarez:2011gd,Alvarez:2020izs,Alvarez:2021zhh,Alvarez:2020qmy,Alvarez:2021qbu}. 
% \cite{Jinno:2020zzs}.

One reason to consider Dirac's conjecture as generally valid is the idea that the quantization of these counterexamples is inconsistent \cite{Henneaux:1992ig}. Here we have shown, however, that the quantization obstruction was not intrinsic to the system, but to the approach of quantization. Noticing that secondary constraints can be treated as initial conditions, there is no need to change the bracket definition, and the system can be quantized in a standard way, preparing the initial state of the wave function accordingly. 

In conclusion, the existence of counterexamples to Dirac's conjecture, along with their consistent quantization, indicates that Dirac's extended dynamics does not hold universally. While it remains a consistent approach, it is important to verify that it does not result in the loss of any crucial degree of freedom in the specific problem that is being handled.

\section*{Acknowledgements}

This work was partially funded by grant FONDECYT 1220862.

\bibliographystyle{unsrt}

\end{document}